# An open question: Are topological arguments helpful in setting initial conditions for transport problems and quantization criteria/ quantum computing for Density Wave physics?

**A.W. Beckwith[1]**

[1] contractor, FNAL/ APS

*abeckwith@UH.edu*

The tunneling Hamiltonian is a proven method to treat particle tunneling between different states represented as wavefunctions in many-body physics. Our problem is how to apply a wave functional formulation of tunneling Hamiltonians to a driven sine-Gordon system. We apply a generalization of the tunneling Hamiltonian to charge density wave (CDW) transport problems in which we consider tunneling between states that are wavefunctionals of a scalar quantum field. We present derived I-E curves that match Zenier curves used to fit data experimentally with wavefunctionals congruent with the false vacuum hypothesis. The open question is whether the coefficients picked in both the wavefunctionals and the magnitude of the coefficients of the driven sine Gordon physical system should be picked by topological charge arguments that in principle appear to assign values that have a tie in with the false vacuum hypothesis first presented by Sidney Coleman. Our supposition is that indeed this is useful and that the topological arguments give evidence as to a first order phase transition which gives credence to the observed and calculated I-E curve as evidence of a quantum switching phenomena in density wave physics, one which we think with further development would have applications to quantum computing, via quantum coherent phase evolution, as outlined in this paper

PAC numbers: 03.75.Lm, 11.27.+d, 71.45.Lr, 75.30.Fv , 85.25.Cp

## INTRODUCTION

This paper's main result has a very strong convergence with the slope of graphs of electron-positron pair production representations. The newly derived results include a threshold electric field explicitly as a starting point without an arbitrary cut off value for the start of the graphed results, thereby improving on both the Zener plots and Lin's generalization of Schwingers 1950 electron-positron nucleation values results for low dimensional systems. The similarities in plot behavior of the current values after the threshold electric field values argue in favor of the Bardeen pinning gap paradigm. We conclude with a discussion of how these results can be conceptually linked to a new scheme of exact evolution of the dynamics of quantum $\phi^4$ field theory in 1+1 dimensions.In several dimensions, we find that the Gaussian wavefunctionals would be given in the form given by Lu. Lu's integration given below is a two dimensional Gaussian wave functional. The analytical result we are working with is a one-dimensional version of a ground-state wave functional of the form

$$|0>^o = N \cdot \exp\left\{-\int_{x,y} \left[(\phi_x - \varphi) \cdot f_{xy} \cdot (\phi_y - \varphi)\right] \cdot dx \cdot dy\right\} \quad (1)$$

Lu's Gaussian wave functional is for a non-perturbed, Hamiltonian as given in Eq. (2) below

$$H_O = \int_x \left[\frac{1}{2} \cdot \Pi_x^2 + \frac{1}{2} \cdot (\partial_x \phi_x)^2 + \frac{1}{2} \cdot \mu^2 \cdot (\phi_x - \varphi)^2 - \frac{1}{2} \cdot I_0(\mu)\right] \cdot dx \cdot dy \qquad (2)$$

We should note that Lu intended the wavefunctional given in Eq. (1) to be a test functional, much as we would do for finding an initial test functional, using a simple Gaussian in computing the ground state energy of a simple Harmonic oscillator variational derivative. calculation. We may obtain a 'ground state' wave functional by taking the one dimensional version of the integrand given in Eq. (1). This means have[12] a robust Gaussian. Lu' Gaussian wave functional set

$$\frac{\partial^2 \cdot V_E}{\partial \cdot \phi^a \cdot \partial \cdot \phi^b} \propto f_{xy} \qquad (3)$$

Here, we call $V_E$ a (Euclidian-time style) potential, with subscripts $a$ and $b$ referring to dimensionality; and $\phi_x$ an 'x dimension contribution' of alternations of 'average' phase $\varphi$, as well as $\phi_y$ an 'y dimension contribution' of alternations of 'average' phase $\varphi$. This average phase is identified in the problem we are analyzing as $\phi_C$

This leads to writing the new Gaussian wavefunctional to be looking like

$$\Psi \equiv c \cdot \exp(-\alpha \cdot \int dx [\phi - \phi_C]^2) \qquad (5)$$

Making this step from Eq. (1) to Eq. (3) involves recognizing, when we go to one-dimension, that we look at a washboard potential with pinning energy contribution from $D \cdot \omega_P^2$ in one- dimensional CDW systems

$$\frac{1}{2} \cdot D \cdot \omega_P^2 \cdot (1 - \cos \phi) \approx \frac{1}{2} \cdot D \cdot \omega_P^2 \cdot \left(\frac{\phi^2}{2} - \frac{\phi^4}{24}\right) \qquad (6)$$

The fourth-order phase term is relatively small, so we look instead at contributions from the quadratic term and treat the fourth order term as a small perturbing contribution to get our one dimensional CDW potential, for lowest order, to roughly look like Eq. (5). In addition, we should note that the c is due to an error functional-norming procedure, discussed below; $\alpha$ is proportional to one over the length of distance between instaton centers

Figure 1 below represents the constituent components of a S-S' pair; the phase value, $\phi_C$, is set to represent a configuration of phase in which the system evolves to/from in the course of the S-S' pair evolution. This leads to

$$c_1 \cdot \exp\left(-\alpha_1 \cdot \int d\tilde{x} [\phi_F]^2\right) \cong \Psi_{initial} \qquad (7)$$

As well as

$$c_2 \cdot \exp\left(-\alpha_2 \cdot \int d\tilde{x} [\phi_T]^2\right) \cong \Psi_{final} \qquad (8)$$

$$f_{xy} \xrightarrow{reduction-to-one-\dim} \delta(x - y)/L^{1+\delta+} \qquad (9)$$

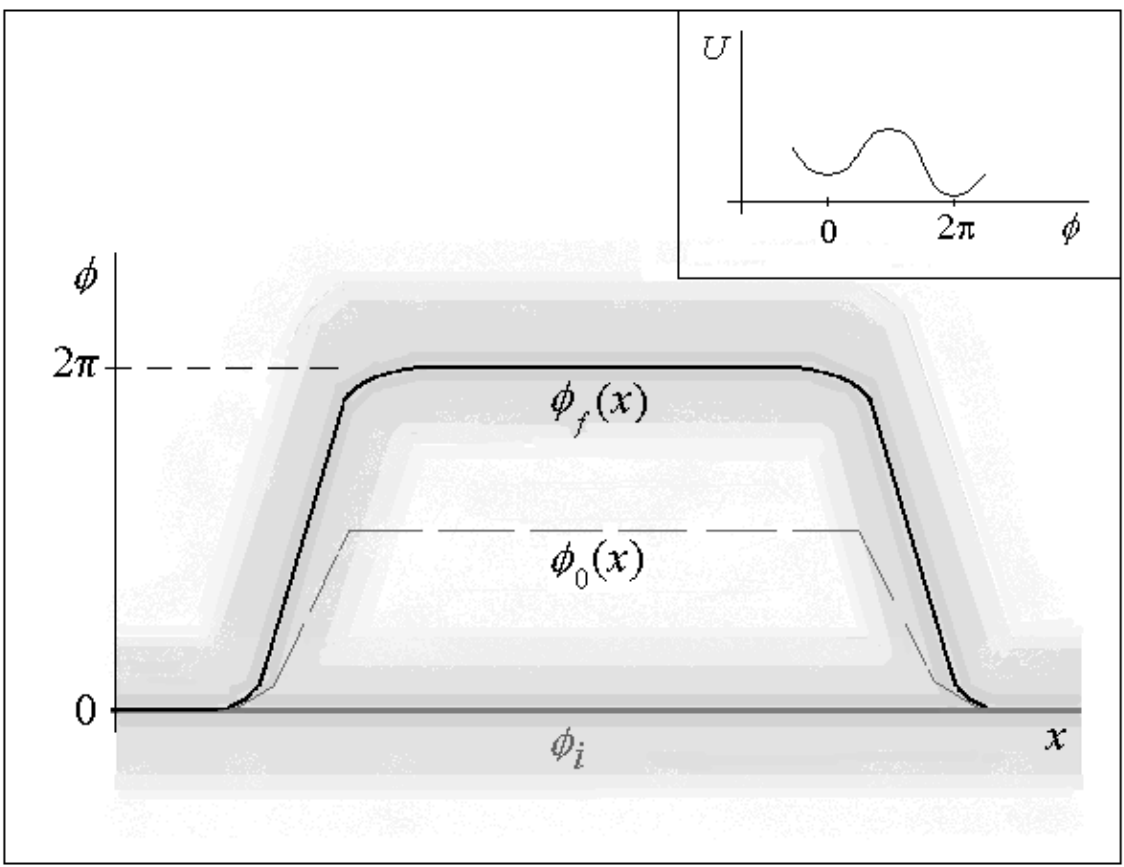

**Fig. 1:** *Evolution from an initial state $\Psi_i[\phi]$ to a final state $\Psi_f[\phi]$ for a double-well potential (inset) in a 1-D model, showing a kink-antikink pair bounding the nucleated bubble of true vacuum. The shading illustrates quantum fluctuations about the initial and final optimum configurations of the field, while $\phi_0(x)$ represents an intermediate field configuration inside the tunnel barrier. The upper right hand side of this figure is how the fate of the false vacuum hypothesis gives a difference in energy between false and true potential vacuum values*

Whereas in multi dimensional treatments, we have

$$f_{xy} \approx \frac{\partial V_{eff}}{\partial r} \tag{10}$$

In the current vs. applied electric field derivation results, we identify the $\Psi_i[\phi]$ as the initial wave function at the left side of a barrier and $\Psi_f[\phi]$ as the final wave function at the right side of a barrier. This can most easily be seen in the following diagram of how the S-S' pair structure arose in the first place, as given by Fig. 2:

# CDW and its Solitons

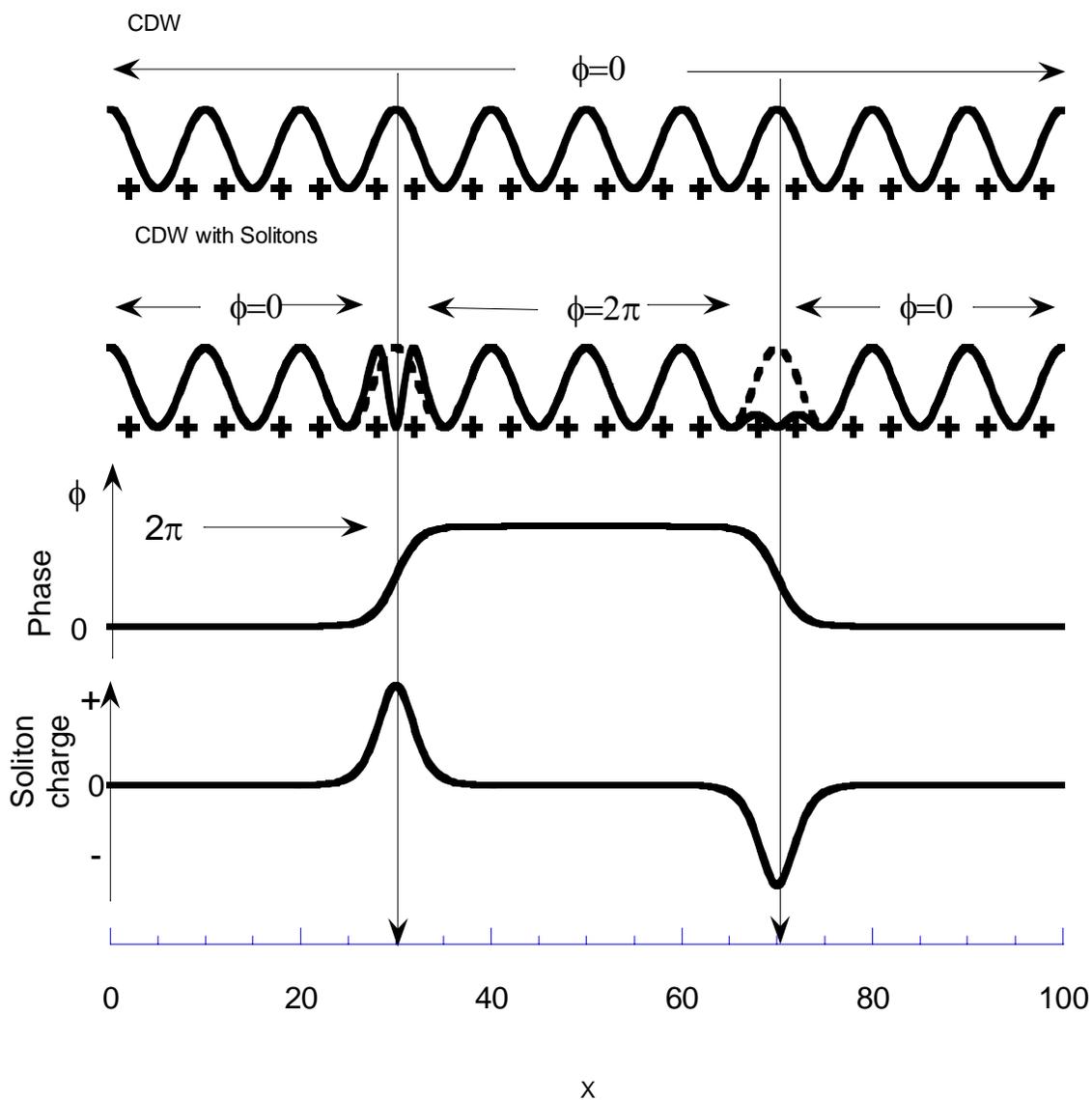

**Fig 2** : *The above figures represents the formation of soliton-anti soliton pairs along a 'chain'. The evolution of phase is spatially given by* $\phi(x) = \pi \cdot [\tanh b(x - x_a) + \tanh b(x_b - x)]$

The tunneling Hamiltonian incorporates wavefunctionals whose Gaussian shape keeps much of the structure as represented by Fig. 2. Following the false vacuum hypothesis, we have a false vacuum phase value $\phi_F \equiv <\phi>_1 \cong$ *very small value*, as well as having in CDW, a final true vacuum $\phi_T \cong \phi_{2\pi} \equiv 2 \cdot \pi + \varepsilon^+$. This led to Gaussian wavefunctionals with a simplified structure. For experimental reasons, we need to have (if we set the charge equal to unity, dimensionally speaking)

$$\alpha \approx L^{-1} \equiv \Delta E_{gap} \equiv V_E(\phi_F) - V_E(\phi_T) \tag{10}$$

This is equivalent to the situation as represented by Fig. 3

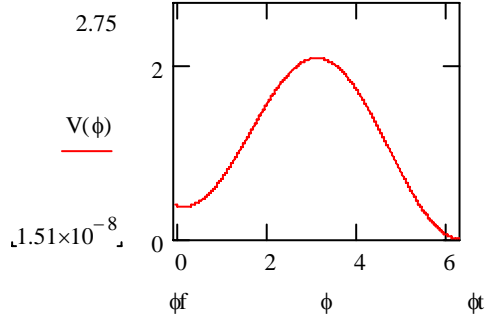

**Fig 3**: *Fate of the false vacuum representation of what happens in CDW. This shows how we have a difference in energy between false and true vacuum values. This eventually leads to a current along the lines of*

$$I \propto \tilde{C}_1 \cdot \left[ \cosh\left[ \sqrt{\frac{2 \cdot E}{E_T \cdot c_V}} - \sqrt{\frac{E_T \cdot c_V}{E}} \right] \right] \cdot \exp\left( -\frac{E_T \cdot c_V}{E} \right) \tag{11}$$

The current expression is a great improvement upon the phenomenological Zener current expression, where $G_P$ is the limiting Charge Density Wave (CDW) conductance.

$$I \propto G_P \cdot (E - E_T) \cdot \exp\left( -\frac{E_T}{E} \right) \text{ if } E > E_T \tag{12}$$

$\qquad$ 0 $\qquad\qquad$ *otherwise*

Fig. 4 illustrates to how the pinning gap calculation improve upon a phenomenological curve fitting result used to match experimental data. The most important feature here is that the theoretical equation takes care of the null values before thre threshold is reached by itself. I.e. we do not need to set it to zero as is done arbitrarily in Eqn (12). This is in any case tied in with a tilted zenier

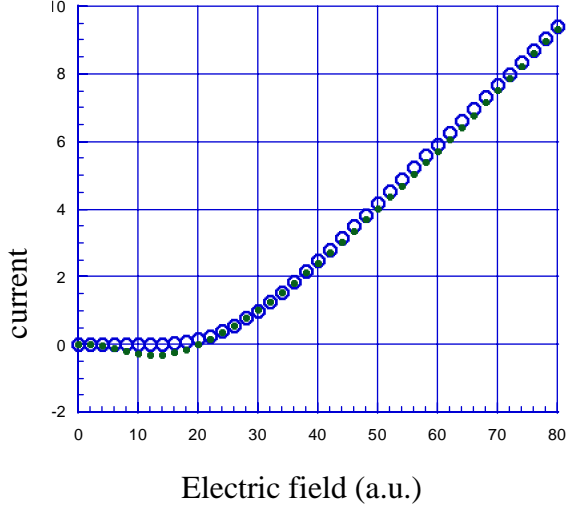

**Fig 4** *Experimental and theoretical predictions of current values versus applied electric field. The dots represent a Zenier curve fitting polynomial, whereas the blue circles are for the S-S' transport expression derived with a field theoretic version of a tunneling Hamiltonian.*

So then, we have $L \propto E^{-1}$.. When we consider a Zener diagram of CDW electrons with tunneling only happening when $e^* \cdot E \cdot L > \varepsilon_G$ where $e^*$ is the effective charge of each condensed electron and $\varepsilon_G$ being a pinning gap energy, we find, assuming that x is the de facto distance between an instanton pair and a measuring device.

In the current vs. applied electric field derivation results, we identify the $\Psi_i[\phi]$ as the initial wave function at the left side of a barrier and $\Psi_f[\phi]$ as the final wave function at the right side of a barrier. Note that Tekman[5] extended the tunneling Hamiltonian method to encompass more complicated geometries. We notice that when the matrix elements $T_{kq}$ are small, we calculate the current through the barrier using linear response theory. This may be used to describe coherent Josephson-like tunneling of either Cooper pairs of electrons or boson-like particles, such as super fluid He atoms. In this case, the supercurrent is linear with the effective matrix element for transferring a pair of electrons or transferring a single boson, as shown rather elegantly in Feynman's derivation of the Josephson current-phase relation. This means a current density proportional to $|T|$ rather than $|T|^2$ since tunneling, in this case, would involve coherent transfer of individual (first-order) bosons rather than pairs of fermions. Note that the initial and final wave functional states were in conjunction with a pinning gap formulation of a variation of typical band calculation structures. This also lead us to after much work to make the following scaling rule which showed up in the linkage between the tilted zenier band model and the distance to a measuring device, which we call x, representing the distance between a modeled instanton structure in density wave physics and a measuring device.

$$\frac{L}{x} \cong c_v \cdot \frac{E_T}{E} \qquad (13)$$

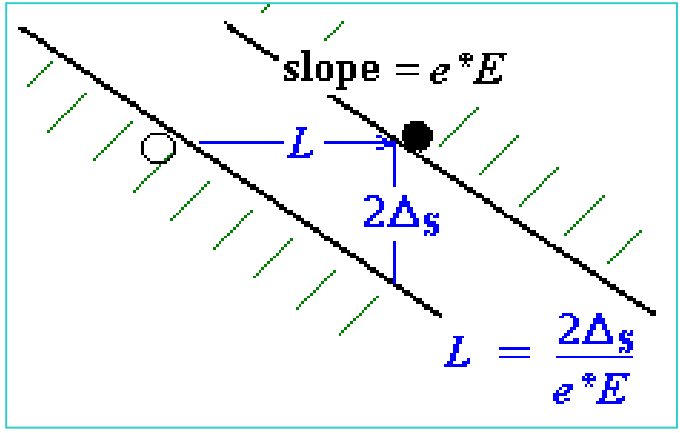

**Fig 5**: *This is a representation of 'Zener' tunneling through pinning gap with band structure tilted by applied E field*

## COMPARISON WITH LIN'S GENERALIZATION

In a 1999, Qiong-gui Lin proposed a general rule regarding the probability of electron-positron pair creation in D+1 dimensions, with D varying from one to three, leading, in the case of a pure electric field, to

$$w_E = (1 + \delta_{d3}) \cdot \frac{|e \cdot E|^{(D+1)/2}}{(2 \cdot \pi)^D} \cdot \sum_{n=1}^{\infty} \frac{1}{n^{(D+1)/2}} \cdot \exp\left(-\frac{n \cdot \pi \cdot m^2}{|e \cdot E|}\right) \qquad (14)$$

When D is set equal to three, we get (after setting $e^2, m \equiv 1$)

$$wIII(E) = \frac{|E|^2}{(4 \cdot \pi^3)} \cdot \sum_{n=1}^{\infty} \frac{1}{n^2} \cdot \exp\left(-\frac{n \cdot \pi}{|E|}\right) \qquad (15)$$

which, if graphed gives a comparatively flattened curve compared w.r.t. to what we get when D is set equal to one (after setting $e^2, m \equiv 1$)

$$wI(E) = \frac{|E|^1}{(2 \cdot \pi^1)} \cdot \sum_{n=1}^{\infty} \frac{1}{n^1} \cdot \exp\left(-\frac{n \cdot \pi}{|E|}\right) \equiv -\frac{|E|}{2 \cdot \pi} \cdot \ln\left[1 - \exp\left(-\frac{\pi}{E}\right)\right] \qquad (16)$$

which is far more linear in behavior for an e field varying from zero to a small numerical value. We see these two graphs in Fig. 6.

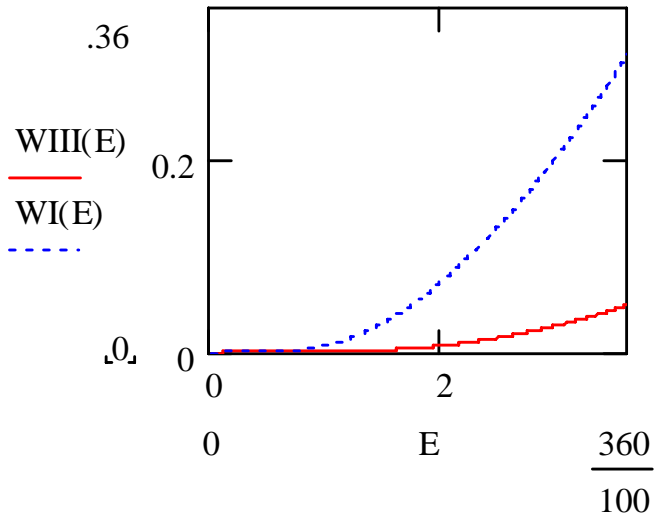

**Fig 6** *Two curves representing probabilities of the nucleation of an electron-positron pair in a vacuum. $wI(E)$ is a nearly-linear curve representing a 1+1 dimensional system, whereas the second curve is for a 3 + 1 dimensional physical system and is far less linear*

This is indicating that, as dimensionality drops, we have a steady progression toward linearity. The three-dimensional result given by Lin is merely the Swinger result observed in the 1950s. When I have $D = 1$ and obtain behavior very similar to the analysis completed for the S-S' current argument just presented, the main difference is in a threshold electric field that is cleanly represented by our graphical analysis. This is a major improvement in the prior curve fitting exercised used in 1985 to curve-fit data.

# CONCLUSION- AND LINKAGE TO EXACT DYNAMICS OF $\phi^4$ FIELD THEORY IN 1+1 DIMENSIONS

We restrict this analysis to ultra fast transitions of CDW; this is realistic and in sync with how the wavefunctionals used are formed in part by the fate of the false vacuum hypothesis.
Additionally, we explore the remarkable similarities between what we have presented here and Lin's expansion of Schwinger's physically significant work in electron-positron pair production. That is, the pinning wall interpretation of tunneling for CDW permits construction of *I-E* curves that match experimental data sets; beforehand these were merely Zener curve fitting polynomial constructions.

Having obtained the *I-E* curve similar to Lin's results gives credence to a pinning gap analysis of CDW transport, with the main difference lying in the new results giving a definitive threshold field effect, whereas both the Zenier curve fit polynomial and Lin's results are not with a specifically delineated threshold electric field. v The derived result does not have the arbitrary zero value cut off specified for current values below given by Miller et al[1] in 1985 but gives this as a result of an analytical derivation. This assumes that in such a situation that the electric field is below a given threshold value.

We used the absorption of a Peierls gap $\Delta'$ term as clearly demonstrated in a numerical simulation paper I wrote to help form solitons (anti-solitons) used in my Gaussian wave functionals for the reasons stated in my. IJMPB article. This is new physics which deserves serious further investigation. It links our formalism formally with a JJ (Josephon junction) approach, and provides analogies worth pursuing in a laboratory environment. The also stunning development is that the plotting of Eqn (16) ties in with the electron-positron plots as given by Lin in Fig (6) for low dimensional systems, which conveniently fits with identification of a S-S' pair with different 'charge centers'

How does this relate to what we can think of concerning questions we raise about phase transitions ?

**Question 1)** Can we use a topological model for phase transitions concerning the change from a false vaccum to a real vacuum? The answer to this one lies in our interpretation of $\alpha \approx L^{-1} \equiv \Delta E_{gap} \equiv V_E(\phi_F) - V_E(\phi_T)$, and in the congruency of Eqn. (16) with the I-E plot given by Eqn. (11). If there is a close analytical match up between Eqn. (11) and Eqn. (16), the topological match up is probably not necessary, and we can de facto stick with analytical derivation of current with its direct match up with electron – positron models. If we cannot get a close to exact congruency between Eqn. (11) and (16) then perhaps we should think of the Bogomolnyi inequality approach mentioned in several publications by this author to explain the inherent breaking of symmetry presented above

**Question 2)** What about linkage to quantum switching devices implied by the abrupt I-E curve turn on implied by **Fig 4** above ? My intuition is that this can be answered if or not there is a phase transition. I.e. if the Bogomolnyi inequality can to be used directly to get $\alpha \approx L^{-1} \equiv \Delta E_{gap} \equiv V_E(\phi_F) - V_E(\phi_T)$ exactly configured as a bridge of the $\phi^4$ field theory in 1+1 dimension to Gaussian wave functionals, we will have quantum switching and ultra fast data transitions along requirements needed for phase transitions. Please see the following references as to how this is modeled

**Question 3)** What about direct analogies as to how we can solve a $\phi^4$ field theory in 1+1 dimensions exactly in terms of real time evolution, without invoking the Bogomolonyi inequality ? This is in terms of taking a real time evolution of the phase is a way to fill in detail alluded to in **Fig 1**, making use of the following

$$\Psi_f[\phi(\mathbf{x})]\bigg|_{\phi \equiv \phi_{Cf}} =$$
$$c_f \cdot \exp\left\{-\int d\mathbf{x}\, \alpha \left[\phi_{Cf}(\mathbf{x}) - \phi_0(\mathbf{x})\right]^2\right\} \rightarrow \qquad (17a)$$
$$c_2 \cdot \exp\left(-\alpha_2 \cdot \int d\tilde{x}\, [\phi_T]^2\right) \cong \Psi_{final},$$

and

$$\Psi_i[\phi(\mathbf{x})]\bigg|_{\phi \equiv \phi_{Ci}}$$
$$= c_i \cdot \exp\left\{-\alpha \int d\mathbf{x}\, [\phi_{ci}(\mathbf{x}) - \phi_0]^2\right\} \rightarrow \qquad (17b)$$
$$c_1 \cdot \exp\left(-\alpha_1 \cdot \int d\tilde{x}\, [\phi_F]^2\right) \equiv \Psi_{initial},$$

This will represent a kink, anti kink combination with the kink given to us as part of Coopers presentation of a sympletic algorithm of updating the operator equations of quantum evolution of a potential system he writes as

$$V[\phi] = -\frac{m^2}{2}\cdot\phi^2 + \frac{\lambda}{4}\cdot\phi^4 \leftrightarrow \frac{1}{2}\cdot D\cdot\omega_P^2\cdot(1-\cos\phi) \approx \frac{1}{2}\cdot D\cdot\omega_P^2\cdot\left(\frac{\phi^2}{2}-\frac{\phi^4}{24}\right) \quad (18)$$

which has a kink solution of the form $\phi[x] = \frac{m}{\sqrt{\lambda}}\cdot\tanh\frac{mx}{\sqrt{2}}$. A kink-anti kink structure so implied by the Gaussian wave functional is stated by Cooper, quoting Moncrief to have an evolution given by a sympletic evolution equation, as given below assuming an averaging procedure we can write as

$$y_i \sim \int_{V_i} dx\,\phi[x,t]/\Delta V_i \approx \text{average of } \phi[x,t] \text{ in a ball about } x_i \text{ of volume } \Delta V_i \quad (19a)$$

And

$$\frac{dy_i}{dt} \equiv \pi_i[t] \quad (19b)$$

And

$$\frac{d\pi_i}{dt} \equiv \frac{1}{a^2}\cdot[y_{i+1} + y_{i-1} - 2y_i] - \lambda y_i^3 + m^2 y_i = F[y_i] \quad (19c)$$

This is assuming that we spatially discretize a Hamiltonian density via

$$\int dx \to a\cdot\sum_i \quad (19d)$$

Following a field theory replacement of $\hat{x} \to \phi_{op}[x,t]$, and a discretized time structure given by $t = j \in$

This leads to the possibility of looking at a quantum foam evolution as given in **Fig 1** via the following sympletic structures, with i the 'spatial component along a chain', and j the 'time component' along a chain. Eqn. (19e) and Eqn (19f) are materially no different than having energy course through a wave lattice as seen in ocean swells accommodating an energy pulse through the water.

$$y_i[j+1] = y_i[j] + \in\cdot\pi_i[j] + \frac{\in^2}{2}\cdot F(y_i[j]) \quad (19e)$$

$$\pi_i[j+1] = \pi_i[j] + \frac{\in}{2}\cdot(F(y_i[j]) + F(y_{i+1}[j])) \quad (19f)$$

A proper understanding of this evolution dynamic should permit a more mature quantum foam interpretation of false vacuum nucleation. This is, of course independent of the datum that adjacent chains in themselves interacting are a necessary condition for the formation of instatons in the first place, as given by the following appendix entry presentation as to a necessary condition for the formation of a kink ( anti kink ) in the first place

## APPENDIX: FORMATION OF THE INSTANTON VIA ADJACENT CHAINS

**What role does the multi chain argument play as far as formulation of the soliton – instanton ? Why add in the Pierls gap term in the first place?**

**Answer**

First of all, we add the following term, based upon the Pierls gap to an analysis of how an instanton evolves

$$H = \sum_n\left[\frac{\Pi_n^2}{2\cdot D_1} + E_1[1-\cos\phi_n] + E_2(\phi_n - \Theta)^2 + \Delta'\cdot[1-\cos(\phi_n - \phi_{n-1})]\right] \quad (1)$$

$$\Pi_n = (h/i) \cdot \partial/\partial\dot\phi_n$$ which then permits us to write

$$U \approx E_1 \cdot \sum_{l=0}^{n+1}[1-\cos\phi_l] + \frac{\Delta'}{2} \cdot \sum_{l=0}^{n}(\phi_{l+1}-\phi_l)^2 \qquad (2)$$

which allowed using $L = T - U$ a Lagrangian based differential equation of

$$\ddot\phi_i - \omega_0^2[(\phi_{i+1}-\phi_i)-(\phi_i-\phi_{i-1})] + \omega_1^2 \sin\phi_i = 0 \qquad (3)$$

with

$$\omega_0^2 = \frac{\Delta'}{m_{e^-} l^2} \qquad (4)$$

$$\omega_1^2 = \frac{E_1}{m_{e^-} l^2} \qquad (5)$$

where we assume the chain of pendulums, each of length $l$, leads to a kinetic energy

$$T = \frac{1}{2} \cdot m_{e^-} l^2 \cdot \sum_{j=0}^{n+1} \dot\phi_j^2 \qquad (6)$$

To get this, we make the following approximation. This has

$$\Delta'(1-\cos[\phi_n - \phi_{n-1}]) \to \frac{\Delta'}{2} \cdot [\phi_n - \phi_{n-1}]^2 + \text{very small H.O.T.s.} \qquad (7)$$

and then consider a nearest neighbor interaction behavior via

$$V_{n.n.}(\phi) \approx E_1[1-\cos\phi_n] + E_2(\phi_n - \Theta)^2 + \frac{\Delta'}{2} \cdot (\phi_n - \phi_{n-1})^2 \qquad (8)$$

Here, we set $\Delta' \gg E_1 \gg E_2$, so then this is leading to a dimensionless Sine–Gordon equation we write as

$$\frac{\partial^2 \phi(z,\tau)}{\partial \tau^2} - \frac{\partial^2 \phi(z,\tau)}{\partial z^2} + \sin\phi(z,\tau) = 0 \qquad (9)$$

**Punch line. Without the Pierls term added in, we do not get a Sine Gordon equation. No Instanton formulation.**